\documentclass[prl,twocolumn,aps,superscriptaddress,footinbib]{revtex4-2}
\usepackage{xspace,amsmath,amsfonts,amssymb,amsbsy}
\usepackage{ntheorem}
\usepackage{graphicx,epstopdf}
\usepackage{bbm}
\usepackage{footnote}
\usepackage[usenames,dvipsnames]{xcolor}
\usepackage{soul}
\usepackage{ulem}

\newcommand\myshade{30}
\colorlet{mylinkcolor}{red}
\colorlet{mycitecolor}{orange}
\colorlet{myurlcolor}{orange}

\usepackage[%
bookmarks=true,
colorlinks,
linkcolor=mylinkcolor!\myshade!black,
urlcolor=myurlcolor!\myshade!black,
citecolor=mycitecolor!\myshade!black,
plainpages=false,
pdfpagelabels,
final,
breaklinks=true
]{hyperref}

\usepackage{physics}
\usepackage{braket}

\begin{document}
\title{
Hamiltonian Engineering of collective XYZ spin models in an optical cavity}%: From one-axis twisting to  two-axis counter twisting models}

\begin{abstract}

Quantum simulation using synthetic quantum systems offers unique opportunities to explore open questions in many-body physics and a path for the generation of useful entangled states. Nevertheless, so far many quantum simulators have been fundamentally limited in the models they can mimic.  Here, we are able to realize an all-to-all interaction with arbitrary quadratic Hamiltonian or effectively an infinite range tunable Heisenberg XYZ model.  This is accomplished by engineering cavity-mediated four-photon interactions between 700 rubidium atoms in which we harness a pair of momentum states as the effective pseudo spin or qubit degree of freedom.  Using this capability we realize for the first time the so-called two-axis counter-twisting model at the mean-field level.
%\clsout{, an iconic XYZ collective spin model that can generate spin-squeezed states that saturate the  Heisenberg limit bound} 
The versatility of our platform to include more than two relevant momentum states, combined with the flexibility of the simulated Hamiltonians by adding cavity tones opens rich opportunities for quantum simulation and quantum sensing in matter-wave interferometers and other quantum sensors such as optical clocks and magnetometers. 
\end{abstract}

\author{Chengyi Luo} \thanks{These authors contributed equally.}%\textsuperscript{1$\,*$}}
\author{Haoqing Zhang} \thanks{These authors contributed equally.}%\textsuperscript{1$\,*$}}
\author{Anjun Chu}
\author{Chitose Maruko}
\author{Ana Maria Rey}
\author{James K.~Thompson}
\email{jkt@jila.colorado.edu}

\affiliation{JILA, NIST, and Department of Physics, University of Colorado, Boulder, CO, USA.}

\date{ \today }

\maketitle

\section{Introduction}
The ability to create and control different many-body interactions is key for entanglement generation, optimization, quantum sensing, and quantum simulation.  Long-range interacting systems have recently shown much promise for the engineering of Hamiltonians that generate interesting correlations that  propagate across the system. Several experimental platforms are making rapid progress, including Rydberg atoms \cite{Browaeys2020,Weidemuller2021Floquet,Wu2021}, polar molecules \cite{Bohn2017,Ye2023Itinerant}, trapped ions \cite{Monroe2017TimeCrystal,Monroe2021}, cavity QED systems \cite{Mivehvar2021} and defect centers in solids \cite{Lukin2020RobustEngineering, Munns2023}. Short range contact interactions in ultra-cold atomic systems are another promising approach \cite{Bloch2008,Schafer2020,Gross2017}, however, it remains an open challenge to reach the sufficiently low-temperatures.

So far, most efforts to engineer Hamiltonians have been limited to XXZ spin models or models that feature both exchange and Ising interactions. Common to these models is the fact that the total magnetization of the spin ensemble is preserved.   However, limited progress has been achieved in engineering more  general spin models, such as XYZ models, which can break both SU(2),  and U(1) symmetries and lead to more general ground-state and out-of-equilibrium many-body behaviors. A few exceptions include experiments in Rydberg atoms using  two-color dressing  \cite{Gross2023RydbergXYZ} with  dynamics limited to pairs of atoms, or   Floquet engineering in  disordered arrays \cite{Weidemuller2021Floquet}. 

Here we experimentally show that photon-mediated interactions between atoms inside an optical cavity can realize a tunable all-to-all Heisenberg XYZ Hamiltonian, a model that offers rich  avenues    for the exploration of  many-body physics and  quantum metrology.
%\clsout{that provide a complementary approach to cold gases relying on contact interactions, with the added benefit of not requiring us to prepare extremely low entropy quantum degenerate gases.} 
We demonstrate the tunability of the XYZ Hamiltonian by explicitly mapping out the evolution of Bloch vectors on the Bloch sphere at the mean-field level for several control parameters that realize different canonical spin Hamiltonians.

%XYZ Hamiltonians can in fact provide important gains in the context of quantum sensing and metrology applications. For instance, the so called two-axis counter-twisting (TACT)~\cite{KitagawaUedaOAT} model is one type of  XYZ model well known  for its capability to generate entanglement  exponentially fast all the way to the Heisenberg bound.  
Collective XYZ Hamiltonians can in fact provide important gains in the context of quantum sensing and simulation applications. For instance, the so called two-axis counter-twisting (TACT) model features non-trivial and exponentially fast phase-space dynamics as compared to SU(2) and U(1) preserving models such as spin exchange or Ising interactions~\cite{leroux2010implementation, Esslinger2012RotonScience, norcia2018cavity,Schleier-Smith2020GapProtection}.   The TACT model was originally proposed more than 30 years ago for generating squeezed states~\cite{KitagawaUedaOAT}, as well as a simpler version,  the so-called one-axis twisting (OAT) or collective Ising interaction~\cite{shirasaki1990squeezing, KitagawaUedaOAT}. Only the TACT model is predicted to produce squeezed states approaching the Heisenberg limit. However, there has been no demonstration of genuine TACT in any platforms due to the challenge of how to realize this more complex non-linear Hamiltonian. Up to date only the OAT model has been realized in experiments including 
trapped ions \cite{Monroe2000EntanglingFourIons, Blatt2008IonOAT}, Bose-Einstein Condensates  \cite{OberthalerSqueezedBECInt, Treutlein2010Squeezing}, atomic cavity-QED \cite{leroux2010implementation, VladanEntangledClockLifetime2010,norcia2018cavity,Schleier-Smith2020GapProtection}, superconducting qubits \cite{Zhu2019SCqubits} and optical interferometers \cite{AndrianovLeuchs2023KerrSqueezing}.   Experiments have  so far only approximated TACT dynamics locally on the Bloch sphere by combining OAT with a transverse drive (the so-called Lipkin-Meshkov-Glick (LMG) model) \cite{Oberthaler2015TaT, Jessen2007CsTaT,li2023improving} or equivalent approximations in spin-nematic dynamics in higher spin systems \cite{ChapmanSqueezing2012,You2017TwinFock}. Here, as a particular case of the Heisenberg XYZ model, we are able to experimentally realize the TACT model using only two dressing lasers. This is a greatly simplified approach as compared to prior theory proposals in atom-cavity systems using four laser tones~\cite{MonikaSorenson2017NJP,zhang2017cavity}.

%Here we experimentally show that photon-mediated interactions between atoms inside an optical cavity can be tuned to realize a tunable all-to-all Heisenberg XYZ Hamiltonian \cl{that provide a complementary approach to cold gases relying on contact interactions, with the added benefit of not requiring us to prepare extremely low entropy quantum degenerate gases.} We demonstrate the tunability of the XYZ Hamiltonian by explicitly mapping out the evolution of Bloch vectors on the Bloch sphere at the mean-field level and witness the TACT dynamics for the first time.
 
The pseudo-spin system here is novel, consisting of two momentum states of atoms freely-falling inside the cavity \cite{Thompson2022SAI,finger2023spin}, making our results of great interest for Bragg matter-wave interferometers \cite{LeePRLBragg1995} that are important for both inertial navigation and fundamental science such as searches for dark matter and  dark energy, detection of gravitational waves, and determination of the fine structure constant \cite{MullerFineStructureAI, GuellatiKhelifaFineStructure2020}. This approach can also be straightforwardly applied to systems with  additional internal levels, making it ideal for developing next-generation quantum-enhanced sensors for technology and exploring a broad range of science from atomic clocks \cite{VuleticSqueezedOpticalClock2020,robinson2022direct} and magnetometers \cite{PolzikSqueezedMagnetometer2010, MitchellSubSQLMagnetometry2010, OberthalerSqueezedMagnetometer2014} to geodesy \cite{LudlowGeodesy2018}.

\begin{figure*}[hbt!]
\centering
\includegraphics[width=0.95 \textwidth]{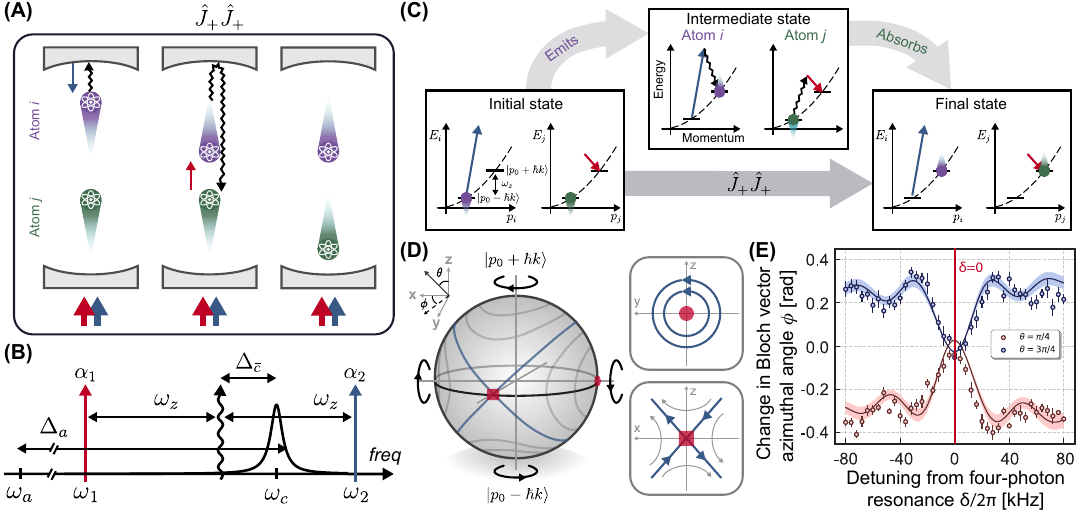}
\caption{\textbf{Experimental overview.} 
    \textbf{(A)} Illustration of the microscopic momentum pair raising process described by the  Hamiltonian $\hat{J}_+\hat{J}_+$, using  two momentum states $p_0\pm\hbar k$ as a  pseudo-spin 1/2 degree of freedom.Initially, two atoms $i$ and $j$ are in the same momentum state along the cavity axis. Dressing lasers are applied to the cavity (red and blue arrows) that allow atom $i$ to absorb a dressing laser photon and emit a photon (squiggly line) into the cavity such that the net photon recoil flips its momentum state by $2\hbar k$. The emitted photon is absorbed by atom $j$, also flipping its momentum state by $2\hbar k$. 
    There also exists the separate momentum exchange process $\hat{J}_+\hat{J}_-$, where atoms initially in opposite momentum states flip their momentum states by emitting and absorbing photons explored in \cite{Thompson2023MomentumExchange}.
   \textbf{(B)} Frequency diagram of the applied dressing lasers with frequencies $\omega_{1,2}$ and coherent state amplitudes $\alpha_{1,2}$.  The emitted photons (squiggly line) are Doppler shifted by $\omega_z$ from the dressing laser frequencies and detuned by $\Delta_c$ from the cavity resonance frequency $\omega_c$.  The cavity is far detuned from the atomic transition frequency $\omega_a$.  \textbf{(C)} Representation of the emission/absorption processes described in (A) but depicted in terms of the atomic  energy versus momentum for atoms $i$ and $j$. The reverse process is also allowed, giving rise to a collective lowering operator described by $\hat{J}_-\hat{J}_-$.
   \textbf{(D)} Dynamics induced by two-axis counter twisting in the form of a Hamiltonian $\hat{J}_x^2-\hat{J}_z^2$ is represented on a Bloch sphere with north and south poles defined by $\ket{p_0 \pm \hbar k}$.
    Top right shows the local circular flows around the stable point at $-\hat{x}$. The local flows for unstable saddle points at $\hat{y}$ with exponential squeezing and anti-squeezing along $\hat{x}\pm \hat{z}$ are shown in bottom right. \textbf{(E)} Observation of the four-photon resonance that generates pair raising and lowering processes.  We scan the dressing laser frequency difference to vary the detuning $\delta$.  With equal dressing laser amplitudes corresponding to realizing the $\hat{J}_x^2$ Hamiltonian when $\delta=0$, we see clear resonances in the observed change of the Bloch vector's azimuthal angle $\phi$ for a Bloch vector prepared near the south pole ($\theta=\pi/4$, red data points) and near the north pole ($\theta=3\pi/4$, blue data points). All error bars reported are $1\sigma$ uncertainties. Simulation results are shown in solid lines with shaded area allowing for 5\% uncertainty in interaction strength. See supplement for details. For all the Hamiltonians engineered later, we will focus on the resonant case with $\delta=0$ as highlighted by the red solid line}. \label{fig:setup}
\end{figure*}

%--------------------------------------------------
\section{Experimental setup}

In the experiment, $^{87}$Rb atoms are laser-cooled inside a vertically-oriented two-mirror standing wave cavity, see Fig.~1\textbf{(A)} and \cite{Thompson2023MomentumExchange, Thompson2022SAI}. A repulsive intra-cavity doughnut dipole trap confines the atoms radially, but allows the atoms to fall along the cavity axis. To prepare atoms in a well-defined momentum state, a pair of laser beams are injected into the cavity to drive velocity-dependent two-photon Raman transitions between ground hyperfine states $\ket{F=1, m_F=0}$ and $\ket{F=2, m_F=0}$. After removing the unselected atoms with a resonant laser push beam, successive microwave pulses are applied to prepare the internal states of the selected atoms in the ground hyperfine state $\ket{F=2,m_F=2}$ (see supplement and \cite{Thompson2023MomentumExchange}).

With about 700 atoms centered at momentum $p_0-\hbar k$, another pair of laser beams is injected along the cavity axis to drive two-photon Bragg transitions connecting the two momentum states $ \ket{p_0\pm \hbar k}$ which defines a two-level spin-1/2 system \cite{Thompson2022SAI, Thompson2023MomentumExchange}.  Here the average momentum is $p_0$, $\hbar$ is the reduced Plank constant, and $k = 2\pi/\lambda$ where $\lambda$ is the wavelength of the Bragg laser beams.  Due to finite momentum spread, one can consider momentum wave packets centered at these two momentum states as considered in detail previously \cite{Thompson2023MomentumExchange}.  Ignoring the finite momentum spread of the selected momentum states, we define $\hat \psi_{\uparrow,\downarrow}^\dagger$ and $\hat\psi_{\uparrow,\downarrow}$ as the operators for creating and annihilating an atom in momentum states $\ket{\uparrow}\equiv\ket{p_0 + \hbar k}$ and $\ket{\downarrow}\equiv\ket{p_0 - \hbar k}$. For mapping to a pseudo-spin model, we define ladder operators $\hat{J}_+ =\hat \psi_\uparrow^\dagger \hat \psi_\downarrow$, $\hat{J}_- = \hat \psi_\downarrow^\dagger \hat\psi_\uparrow$ and spin projection operators $\hat{J}_x = \frac{1}{2} \left( \hat{J}_+ + \hat{J}_- \right) $, $\hat{J}_y = \frac{1}{2i} \left( \hat{J}_+ - \hat{J}_- \right) $ and $\hat{J}_z = \frac{1}{2} \left( \hat\psi_\uparrow^\dagger \hat\psi_\uparrow - \hat \psi_\downarrow^\dagger \hat\psi_\downarrow \right) $. 

%-----------------------------------------------------
\vspace{2mm}
As shown in Fig.~\ref{fig:setup}\textbf{(B)}, the cavity frequency $\omega_c$ is detuned from the atomic transition $\ket{F=2,m_F=2} \rightarrow \ket{F'=3,m_{F'}=3}$ by $\Delta_a =\omega_c-\omega_a = 2\pi \times 500$~MHz, which is much larger than the excited state decay rate $\Gamma=2\pi \times 6~$MHz and the cavity power decay rate $\kappa = 2\pi \times 56(3)~$kHz. A series of Bragg pulses can be applied to realize a Mach-Zehnder matter-wave interferometer (i.e.~$\pi/2 - \pi - \pi/2$), in which the wave packets first separate in position and then re-overlap. When the two wave packets are overlapped, the interference between them forms an atomic density grating with period $\lambda/2$, which matches the standing wave of a cavity mode. 

As the atoms move along the cavity axis, the density grating is periodically aligned to the standing-wave of the cavity mode, leading to modulation of the cavity resonance frequency at the two-photon Doppler frequency $\omega_z = 2 k p_0/m \approx 2\pi \times 500~$kHz, where $m$ is the mass of $^{87}$Rb. To have this modulation to mediate an effective atom-atom interaction, we typically apply two $\sigma^+$ polarized dressing laser tones (see Fig.~\ref{fig:setup}\textbf{(A)}) at frequencies $\omega_{1,2}$ (see Fig.~\ref{fig:setup}\textbf{(B)}), with complex amplitudes  $\alpha_{1,2}$  corresponding to the field that would be established inside the cavity were no atoms in the cavity and implicit units $\sqrt{\mathrm{photons}}$. The atom-induced cavity frequency modulation leads to the generation of modulation sideband tones at frequencies $\omega_{1,2}\pm \omega_z.$  In the following simplifications, we will assume $\omega_2> \omega_1$.

The key insight is that different combinations of the dressing lasers and their atom-induced sideband tones will induce different virtual four-photon processes which will manifest as all-to-all exchange interactions $\hat{J}_+\hat{J}_-$~\cite{Thompson2023MomentumExchange} and pair-raising $\hat{J}_+\hat{J}_+$ (lowering  $\hat{J}_-\hat{J}_-$)  processes  as  shown in Fig.~\ref{fig:setup}\textbf{(A,C)}. After adiabatically eliminating the cavity fields using second order perturbation theory (see \cite{Thompson2023MomentumExchange} and supplement), we obtain an effective time-dependent atom-only Hamiltonian in an appropriate frame rotating at $\omega_z$
\begin{equation}
    \hat{H} = \chi_e \hat{J}_+ \hat{J}_{-} + \left( \chi_{p} e^{i \delta t } \hat{J}_+\hat{J}_+ + \chi_{p}^* e^{-i \delta t } \hat{J}_- \hat{J}_- \right),
\label{eq:Heff}
\end{equation}
with the exchange and pair-raising/lowering couplings given by
\begin{equation}
\begin{split}
    \chi_e & = \left( \frac{g_0^2}{4\Delta_a} \right)^2  \left( \frac{|\alpha_1|^2}{\Delta_{\bar{c}} + \delta/2 } + \frac{ |\alpha_2|^2}{\Delta_{\bar{c}}- \delta/2 } \right) , \\
    \chi_{p} & = \left( \frac{g_0^2}{4\Delta_a} \right)^2 \frac{|\alpha_1 \alpha_2| e^{i \phi_\mathrm{int}}} {2}  \left( \frac{1}{\Delta_{\bar{c}} + \delta/2} +\frac{1}{\Delta_{\bar{c}} - \delta/2} \right) .
\end{split}
\label{eq:chi}
\end{equation}
Here, $\Delta_{\bar{c}}=\left(\omega_2+\omega_1\right)/2 - \omega_c$ is the average detuning of the two dressing lasers from cavity resonance, typically set to be less than 1~MHz. $\delta=\left( \omega_2 - \omega_1 \right)-2\omega_z$ is the detuning from four-photon resonance, $g_0=2\pi \times 0.48$~MHz is the maximal atom-cavity Rabi coupling at an anti-node of the cavity mode, and  $\phi_\mathrm{int}= \arg \left( \alpha_2 \alpha_1^* \right) - \phi_\mathrm{B}$ is the differential phase between the two dressing laser tones relative to the phase of Bragg coupling $\phi_\mathrm{B}$ which forms initial  density grating (see supplement). Collective cavity dissipation is dealt with separately (see supplement).

We will focus on the resonant case $\delta=0$ and $\phi_\mathrm{int}=0$, though we show example data in Fig.~\ref{fig:setup}\textbf{(E)} (see Method) that clearly exhibits a resonance in the interaction-induced dynamics at $\delta=0$ as the pair raising/lowering processes are tuned into and out of resonance by tuning the dressing laser frequency difference.  In the resonant case, the Hamiltonian of Eq.~\eqref{eq:Heff}  reduces to:
\begin{equation}
\begin{aligned}
    \hat{H}
    &=\left(\chi_e+\chi_p\right) \hat{J}^2_x + \left(\chi_e-\chi_p\right) \hat{J}^2_y \\ \label{eq:heff}
    &= \chi_{e} \hat{\bf J}\cdot\hat{\bf J} + \chi_p \left( \hat{J}_{x}^{2} - \hat{J}_{y}^{2} \right) - \chi_{e} \hat{J}_{z}^{2},
\end{aligned}
\end{equation} 
where  we defined the collective angular momentum operator, $\hat{\bf J}=\{\hat{J}_{x},\hat{J}_{y},\hat{J}_{z}\}$ and introduced the  collective  Heisenberg interaction $\hat{\bf J}\cdot\hat{\bf J}$. The latter acts as a constant for any eigenstate of the collective angular momentum operator $\hat{\bf J}\cdot\hat{\bf J}$, such as the collective states with eigenvalue $N/2(N/2+1)$. 
In the presence of single particle inhomogeneities, this term  opens a many-body gap  that help promote spin locking which we explored before for the  protection of coherences against dephasing~\cite{norcia2018cavity,Thompson2023MomentumExchange,Schleier-Smith2020GapProtection}. Here we instead focus on the fully collective dynamics and therefore  without loss of generality we can  add a generic $ \chi_z\hat{\bf J}\cdot\hat{\bf J}$ term without affecting the dynamics.  As such, in our system we are able to engineer dynamics governed by an XYZ Hamiltonian $\hat{H}=\chi_x \hat{J}^2_x + \chi_y \hat{J}^2_y+\chi_ z \hat{J}^2_z$ with interaction strengths $\chi_x=\left(\chi_e+\chi_p+\chi_z\right)$ and $\chi_y=\left(\chi_e-\chi_p+\chi_z\right)$. The XYZ Hamiltonian is highly tunable by simply adjusting the relative power in the two applied dressing lasers since $\chi_e$  scales as $\left|\alpha_2\right|^2$, $\left|\alpha_1\right|^2$ and $\chi_p$ scales as $\left|\alpha_2\alpha_1\right|$.  

\begin{figure*}[hbt!]
\centering
\includegraphics[width=0.95\textwidth]{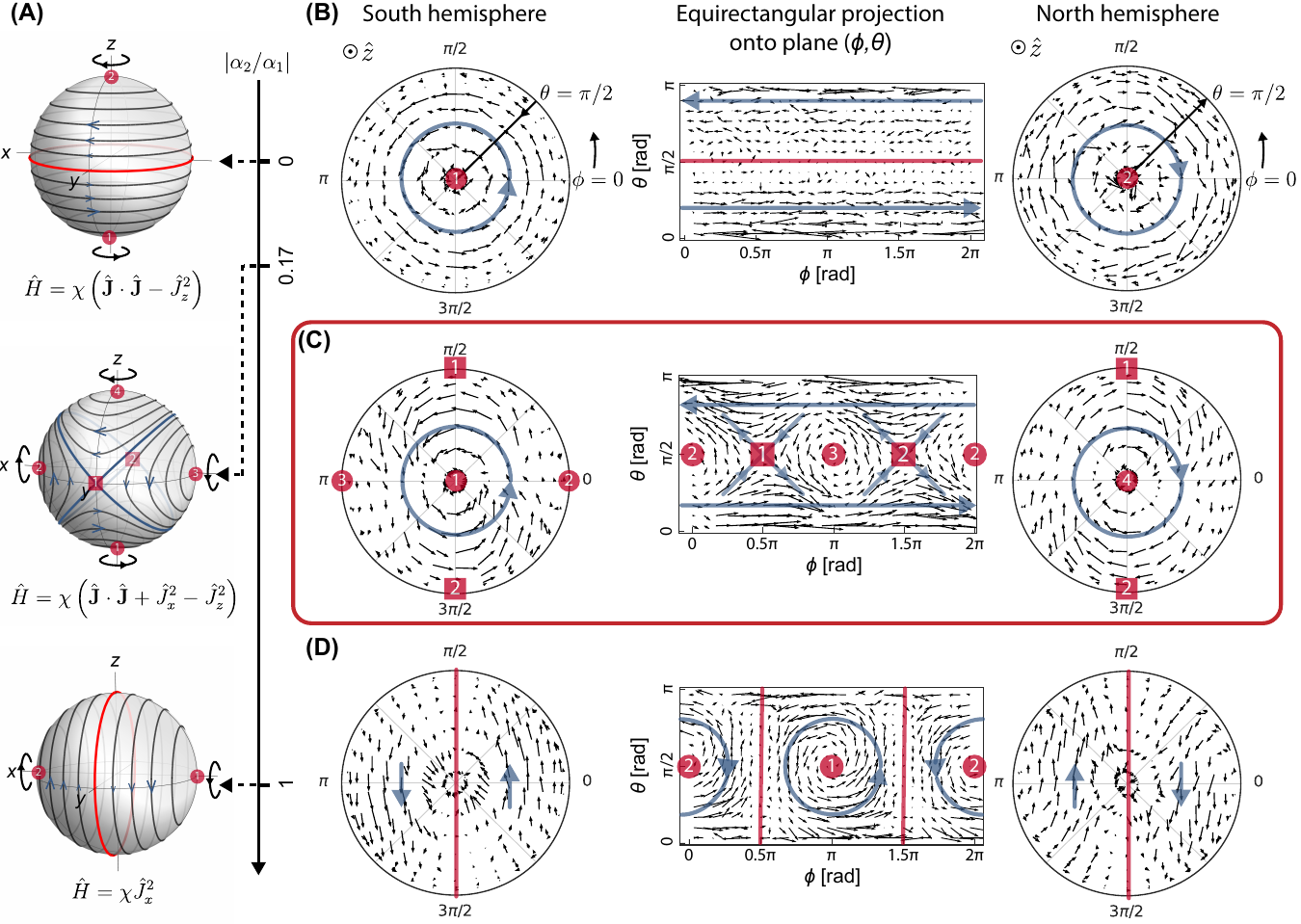}
\caption{\textbf{Evolution under different Hamiltonians} 
    \textbf{(A)} Setting different amplitude ratios between the two dressing tones $|\alpha_2/\alpha_1|$ gives rise to different XYZ Hamiltonians, with specific examples shown here including one-axis twisting $\hat{\bf J}\cdot\hat{\bf J}-\hat{J}_z^2$, two-axis counter twisting $\hat{\bf J}\cdot\hat{\bf J} + \hat{J}_x^2 - \hat{J}_z^2$ and one-axis twisting $\hat{J}_x^2$. The experimentally observed dynamics are shown in corresponding rows in \textbf{(B, C, D)}, where the tail of each vector indicates the initial position of the Bloch vector $\mathbf{J}_i$ on the Bloch sphere, and the arrow indicates the displacement $\mathbf{T}$ after a brief period of evolution under the corresponding Hamiltonians.
   The left (right) panels are for initial Bloch vectors on the south (north) hemisphere. The polar angle $\theta$ of the initial Bloch vector linearly increase from $\pi/2$ at the rim to $\pi$ in the middle for the left plot, whereas $\theta$ decrease from $\pi/2$ at the rim to $0$ in the middle for the right plot. 
   The middle panels are equirectangular projections. 
    In each case, the qualitatively observed stable fixed points are marked with numbered red circles, and the unstable fixed points are marked with numbered red squares. The blue lines are shown to indicate directions of observed flow.  The red lines indicate lines where the dynamics are zero and separate regions of opposite flows on the Bloch sphere.
    }  
\label{fig2}
\end{figure*}

To realize the XYZ Hamiltonians of Fig.~\ref{fig2} and \ref{fig3}, we apply the two dressing laser tones separated by $\Delta_p= \omega_2 -\omega_1 = 2\omega_z \approx 2\pi\times 1$~MHz, and with average dressing laser detuning from the cavity set to $\Delta_{\bar{c}}=2\pi \times 200$~kHz as shown in Fig.~\ref{fig:setup}. Because the atoms are accelerating due to gravity, $\omega_z$ changes linearly in time at a rate $d\omega_z/dt = 2\pi \times 25.11~$kHz/ms.  To compensate for this, we linearly ramp the dressing laser separation at a rate $ d\Delta_p/dt = 2 \,d\omega_z/dt$.  Similarly, the difference frequency of the applied Bragg coupling tones is ramped at  $d \omega_z/dt$.   For the TACT Hamiltonian dynamics presented in Fig.~\ref{fig2} and \ref{fig3}, the total power of the incident dressing lasers was approximately 200~pW or 900 photons/$\mu$s to realize an  interaction strength of  $\chi_p=2\pi \times 1.25$~Hz.

%-----------------------------------------------------
\vspace{2mm}
\section{Mean-field dynamics}
At the mean-field level, we can define the Bloch vector $\mathbf{J}\equiv(J_x,J_y,J_z)=(\langle \hat{J}_{x}\rangle,\langle \hat{J}_{y}\rangle,\langle \hat{J}_z\rangle)$ and approximate the Hamiltonian as $\hat{H}= \mathbf{B}(\mathbf{J})\cdot \hat{\mathbf{J}}$.  In this way,  the collective dynamics are driven by a self-generated effective magnetic field $\mathbf{B}(\mathbf{J})=(2 \chi_x  {J}_{x}, 2 \chi_y {J}_{y},2 \chi_z {J}_{z})$ which depends  on the instantaneous collective spin projections.

We can derive the equations of motion of the collective Bloch vector, from Eq. \eqref{eq:heff} (see Method), which simplify to a nonlinear torque equation $d \mathbf{J} / dt = \mathbf{B} (\mathbf{J}) \times \mathbf{J}  \equiv  \mathbf{T} (\mathbf{J})$. One  can  identify the fixed points  $\mathbf{J}_{\rm fix}$ as the points where $\mathbf{T}(\mathbf{J}_{\rm fix})=0$. 

To understand the dynamics near the fixed points, it is useful to follow a standard stability analysis by diagonalizing the Jacobian matrix $M(\mathbf{J})=\partial \mathbf{T} / \partial \mathbf{J} |_{\mathbf{J}=\mathbf{J}_{\rm fix}}$.   The local motion near these fixed points is illustrated in Fig.~\ref{fig:setup}\textbf{(D)} (right). We use red circles for stable points with purely imaginary eigenvalues. The Bloch vector evolves on stable closed orbits indicated by the blue circular traces.  The red squares are used to denote unstable saddle points with real eigenvalues with opposite signs. The unstable saddle points are labelled by the red squares. The eigenvalues of the Jacobian matrix at the saddle points are real but with opposite signs. In Fig.~\ref{fig:setup}\textbf{(D)} (bottom right), the dynamics show exponential divergence from the origin (indicated by outward blue arrows) along $\hat{x}+\hat{z}$, corresponding to the positive eigenvalue \cite{PoggiDeutsch2023PRXQuantum}. The negative eigenvalues are indicated by the convergence towards the origin (inward blue arrows) along $\hat{x}-\hat{z}$.
%----------------------------------------------------------------
\section{Dynamics on the Bloch sphere}
In the experiment, we probe the local dynamics induced by the above Hamiltonian with various values of $\chi_e$ and $\chi_p$.  To do this, we vary the phase and duration of the Bragg pulse to prepare an initial pseudo-spin coherent state $\mathbf{J}_i$. Before the atomic wave packets separate,  we then apply the interaction for a short time $\Delta t = 50\mu$s satisfying $\chi N \Delta t \ll 1$ and measure the change in azimuthal angle $d\phi$ and polar angle $d\theta$ to obtain the final Bloch vector $\mathbf{J}_f$ after the interaction. This is achieved by repeating the experiment and applying additional appropriate rotations before measuring populations in the two momentum states (see supplement).  The local flow vector is then determined by the torque $\mathbf{T}(\mathbf{J}_i) \approx \Delta\mathbf{J}/\Delta t =\left(\mathbf{J}_f - \mathbf{J}_i\right)/\Delta t$.

In Fig.~\ref{fig2}\textbf{(A)}, we show the predicted flow vectors $\mathbf{T}(\mathbf{J}_i)$ on the Bloch sphere for three example Hamiltonians of interest. Different Hamiltonians are obtained by changing the ratio of the dressing laser amplitudes $\left|\alpha_2/\alpha_1\right| = 0, 0.17$, and $1.0$. 

We show the measured flow vector $\mathbf{T}(\mathbf{J}_i)$ in Fig.~\ref{fig2}\textbf{(B)} with each row aligned to the corresponding example presented in Fig.~\ref{fig2}\textbf{(A)}.  In each case, the flow vector start at $\mathbf{J}_i$ and end at $\mathbf{J}_f$. The left and right panels are polar plots (radial coordinate linear in polar angle) of the dynamics on the south/north hemisphere looking from the north poles of the Bloch sphere. The middle panels are equirectangular projections to show dynamics near the equator. From these vector maps, we can make qualitative comparisons based on the geometry of the flow. For both theoretical and experimental results, the stable fixed points and unstable saddle points labeled as numbered red circles and squares respectively \cite{PoggiDeutsch2023PRXQuantum,Witkowska2015TACTdynamics}.  

\subsection{OAT:  ${\hat{J}_z^2}$ interaction}
In the first or top row of Fig.~\ref{fig2} , we consider the simplest case with $\chi_{p}=0$. This is achieved by turning off one of the dressing lasers ($\left|\alpha_2/\alpha_1\right|=0$), leading to the Hamiltonian $\hat{H}=\chi_{e} \left( \hat{\bf J}\cdot\hat{\bf J} - \hat{J}_z^2\right)$. This Hamiltonian, referred to as the OAT Hamiltonian \cite{KitagawaUedaOAT}, maintains U(1) symmetry, thereby conserving ${J}_z$.
At the mean-field level, it features  a constant effective magnetic field along the $\hat{z}$ direction, which results in the rotation of the collective Bloch vector about the $\hat{z}$-axis at a uniform constant angular frequency,$-2\chi_e {J}_z$. As expected, we observe in Fig.~\ref{fig2}\textbf{(B)} two stable fixed points (red 1 and 2 circles), and a reversal of the circulation across the equator (i.e.~$J_z=0$) where there are no dynamics (red line).  This $J_z$ dependent circulation leads to shearing of the quantum noise in the orientation of a Bloch vector on the Bloch sphere, a semiclassical explanation for how OAT dynamics generate spin-squeezed states \cite{KitagawaUedaOAT,Thompson2022SAI}. We note that while the term
$\hat{\bf J}\cdot\hat{\bf J}$ is trivial for our current observations at short times, at longer times, when inhomogeneities in our system  manifest, it can lead to important dynamical effects as shown in Ref. \cite{Thompson2023MomentumExchange}.

\subsection{OAT:  ${\hat{J}_x^2}$ interaction}
Next, we consider the bottom row in Fig.~\ref{fig2} with $\chi_{p} = \chi_{e} $. This is achieved by using equal dressing laser amplitudes $ \left| \alpha_2 / \alpha_1 \right| = 1$.
Here, the Hamiltonian is  $ \hat{H} = 2\chi_{e}\hat{J}_x^2 $, leading to OAT dynamics along the $\hat{x}$ direction.  The corresponding dynamics are induced at the mean-field level by a magnetic field along $\hat{x}$ that preserves $  {J}_x  $ and induces a rotation about $\hat{x}$  with constant angular frequency $ 4\chi_{e} {J}_{x} $. 
It is noteworthy that the interaction strength here is twice that of the $\hat{J}^2_z $ case, attributed to the use of two dressing laser tones. The data in Fig.~\ref{fig2}\textbf{(D)} qualitatively shows two stable fixed points along $\hat{x}$ labeled by red circles 1 and 2, and a reversal of the sign of circulation across ${J}_{x} $ highlighted by the red lines.

\subsection{TACT:  $\hat{J}_x^2-\hat{J}_z^2$ interaction}
Finally, we come to the case that achieves TACT, as shown in the middle row of Fig.~\ref{fig2}. In this case, the ratio of dressing lasers amplitudes is set to be $|\alpha_2 / \alpha_1|=\left(\sqrt{2}-1\right)/\left(\sqrt{2}+1\right)\approx 0.17$, which produces $\chi_{e} = 3\chi_{p}$. The Hamiltonian then become $\hat{H}=2 \chi_{p} \left( 2\hat{J}_{x}^2 + \hat{J}_{y}^2 \right)=2 \chi_{p}\left( \hat{\bf J}\cdot\hat{\bf J}+ \hat{J}_{x}^2 - \hat{J}_{z}^2\right)$. 

%\cl{The first and the last rows in Fig.~\ref{fig2} both display OAT dynamics oriented along different axes. The resulting flow lines are symmetric under $\pi$ rotation around $\hat{y}$. The flow lines for the TACT model shown in the middle row, break both  SU(2) and U(1) symmetries  (as well as the  OAT's flow lines). The TACT  nevertheless, is  symmetric under a $\pi/2$ rotation along $\hat{y}$ plus a time-reversal operation.}

The corresponding mean-field magnetic field is now $\mathbf{B}(\mathbf{J})=4\chi_p \left({J}_x, 0, -{J}_z  \right)$, where we have again ignored the Heisenberg term since we will only consider dynamics at constant Bloch vector length. The theoretical flow line for the dynamics is depicted in Fig.~\ref{fig2}\textbf{(A)} (middle), showing four stable fixed points along $\pm \hat{x}$ and $\pm \hat{z}$, as well as two unstable fixed points along $\pm \hat{y}$ connected by great circles inclined at $\pm \pi/4$ to the equatorial plane (red circles). 
In comparison to OAT, there exist two twisting fields thus only the two points intersected by two blue circles with ${J}_x={J}_z=0$, exhibit the maximum shearing dynamics, which serve as the unstable saddle points.

%To realize the XYZ Hamiltonians of Fig.~\ref{fig2}, \cl{we apply two dressing laser tones separated by $\Delta_p= \omega_2 -\omega_1 = 2\omega_z \approx 2\pi\times 1$~MHz.  The average dressing laser detuning from the cavity is is set to $\Delta_{\bar{c}}=2\pi \times 200$~kHz as shown in Fig.~\ref{fig:setup}. Because the atoms are accelerating due to gravity, $\omega_z$ changes linearly in time at a rate $d\omega_z/dt = 2\pi \times 25.11~$kHz/ms.  To compensate for this, we linearly ramp the dressing laser separation at a rate $ d\Delta_p/dt = 2 \,d\omega_z/dt$.  Similarly, the difference frequency of the applied Bragg coupling tones is ramped at  $d \omega_z/dt$. }  \cl{For the TACT Hamiltonian dynamics presented in Fig.~\ref{fig2} and \ref{fig3}, the total power of the incident dressing lasers was approximately 900 photons/$\mu$s to realize an  interaction strength of  $\chi_p=2\pi \times 1.25$~Hz.} 

In the left and right panels of Fig.~\ref{fig2}\textbf{(C)}, the observed stable fixed points are labeled with red circles numbered 1 to 4.  The circulations of the flow lines are opposite for stable fixed points on opposite sides of the Bloch sphere. In the middle panel, we observe two unstable points labeled with red squared 1 and 2 in Fig.~\ref{fig2}\textbf{(C)}. The two unstable fixed points have flow lines that either diverge from them (blue arrows outward) or converge towards them  (blue arrows inward), with the two flows orthogonal to each other  (see the discussion below). 
These findings constitute the first direct observation of genuine  TACT dynamics.
We note that for a particular set of parameters of the LMG Hamiltonian, for example when $\hat H=\chi \hat J_z^2+ \delta \hat{J}_y$ with $\delta\sim \chi N$, and when the Bloch vector initially points along the $\hat{y}$ direction~\cite{PoggiDeutsch2023PRXQuantum, Witkowska2015TACTdynamics}, the flow lines can resemble the ones close to a saddle point of the TACT model (see supplement). However, the instability is restricted to this single point, in contrast to the full TACT  which features two independent unstable points.

The observed dynamics reflect the different symmetries of the TACT and OAT Hamiltonians. Both of the observed OAT dynamics are symmetric under $\pi$ rotation around $\hat{y}$ and have continuous rotational symmetry about the twisting axis.  In contrast, the flow lines for the TACT dynamics, shown in the middle row, does not have such symmetries. The TACT dynamics nevertheless, is symmetric under a $\pi/2$ rotation along $\hat{y}$ plus a time-reversal operation.

\begin{figure}[hbt!]
	\centering
	\includegraphics[width=0.5\textwidth]{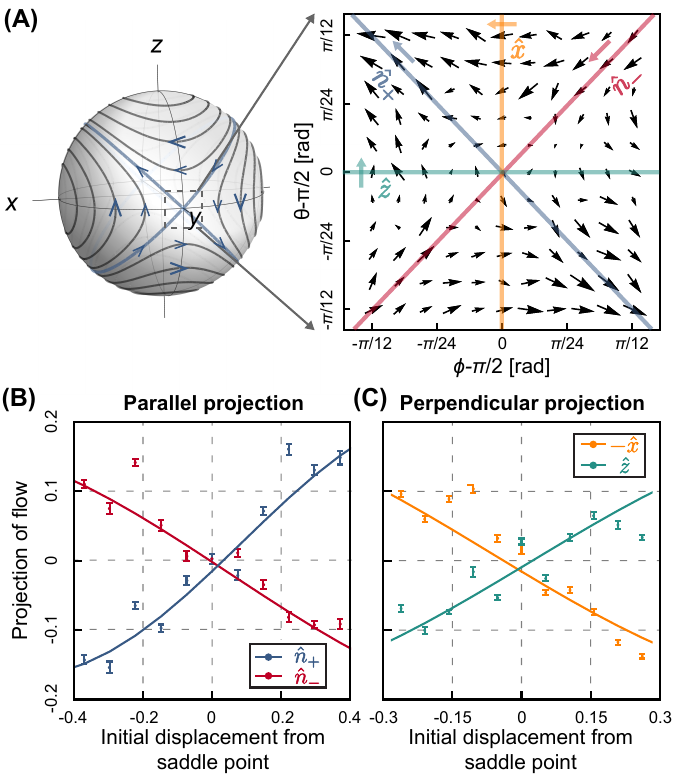}
	\caption{
        \textbf{Dynamics near saddle points.}
	    \textbf{(A)} Measured local flow vector map around the saddle point of the TACT dynamics. 
        \textbf{(B)} Blue and red points are parallel projections of the local flow vector onto the $\hat{n}_{\pm}$ axis $\Delta \mathbf{J} \cdot \hat{n}_\pm/\left(N/2\right)$ as a function of initial displacement $d \mathbf{J}_i \cdot \hat{n}_{\pm}/\left(N/2\right)$ from the saddle point along the blue and red lines in \textbf{(A)}. Blue and red solid lines are simulation results. The linear dependence confirms the predicted exponentially growing / shrinking dynamics along the squeezing / anti-squeezing axes. 
        \textbf{(C)} Green and orange points are the perpendicular projection of the local flow vector as a function of initial angular displacement from the saddle point along the green and orange lines in \textbf{(A)} ($\hat{n}=\hat{x}$ and $
        \hat{z}$) onto $\hat{n}\times\hat{y}=\hat{z}$ and $-\hat{x}$ axis ($\Delta \mathbf{J}_z$ and $- \Delta \mathbf{J}_{x}$).
        % for an initial state at the center of the saddle  point. 
        Orange and green solid lines are simulation results. 
    }
    \label{fig3}
\end{figure}

For a quantitative comparison, we examine the dynamics near the saddle point $\mathbf{J}_{\rm sad}$ along the $\hat{y}$ axis on the Bloch sphere.  In Fig.~\ref{fig3}\textbf{(A)}, we map the displacement $\mathbf{T}\left(\mathbf{J}_i\right)$ as a function of the initial Bloch vector orientation $\mathbf{J}_i = \mathbf{J}_{\rm sad} + d \mathbf{J}_i$. We scan the initial Bloch vector angles $\theta_i$ and $\phi_i$ over a range $\pm \pi/12$ centered about $\pi/2$ (i.e. about $\hat{y}$) with discrete points sampled using a detailed $11\times 11$ grid. %\cl{At the mean-field level,  collective decay (superradiance) can cause a global displacement along $\hat{z}$  of a  Bloch vector initialized  around the saddle point. To compensate for it and remove undesirable dynamics due to superradiance, we subtract  the initial displacement  from every data point for both experimental and theoretical results. The magnitude of the subtracted displacement agrees well with the theoretical predictions (see supplement.)} 
In Fig.~\ref{fig3}, we remove a global displacement along $\hat{z}$ that is common to all data points and which arises from collective or superradiant decay (see supplement.) 

The mean-field equations of motion for the two orthogonal directions $\hat{n}_\pm=\left(\hat{x}\pm\hat{z}\right)/\sqrt{2}$ are:
\begin{equation}
\begin{aligned}
\frac{d }{dt} ({J}_{x}+{J}_{z}) &=4\chi_{p}{J}_{y}( {J}_{x}+{J}_{z}) \\
\frac{d}{dt} ({J}_{x}-{J}_{z}) &=-4\chi_{p}{J}_{y}({J}_{x}-{J}_{z}).
\end{aligned}\label{eq:local}
\end{equation}
For the small range of angles around the $\hat y$ axis sampled in these measurements, one can assume ${J}_y \approx N/2$, and therefore find out that the displacement's time derivative increases linearly with the displacement itself, confirming the predicted dynamics that change exponentially over time.

We first focus on the data points along the two directions $\hat{n}_\pm$ (depicted by blue and red lines in Fig.~\ref{fig3}\textbf{(A)}) and compute the parallel projection of the flow $\Delta \mathbf{J} \cdot \hat{n}_{\pm}/\left(N/2\right)$ as a function of initial displacement from the saddle point $d \mathbf{J}_i \cdot \hat{n}_{\pm}/\left(N/2\right)$, as shown in Fig.~\ref{fig3}\textbf{(B)}. The observed linear relationship between the projection of the flow and initial displacement matches well with the prediction from the simulation results (solid lines) which go beyond the linear approximation by solving the non-linear equations (Eq.~\eqref{eq:local}). The small differences observed between the  $\Delta \mathbf{J} \cdot \hat{n}_{\pm} /\left(N/2\right)$ slopes stem from the finite duration of the interaction, which extends the dynamics beyond the linear response regime.

The unstable dynamics explain why an initial circular distribution centered around the saddle point shears as a function of time by squeezing (anti-squeezing) along the $\hat{n}_-$ ($\hat{n}_+$) direction at a rate exponentially faster than the linear growth seen in OAT.  This combined with the global (over the full Bloch sphere) dynamical behavior allows the TACT to directly approach the fundamental Heisenberg limit on phase estimation \cite{KitagawaUedaOAT,PoggiDeutsch2023PRXQuantum}. 

% In Fig.~\ref{fig3}\textbf{(C)}, we examine the perpendicular projection of the interaction-induced displacements $\Delta \mathbf{J}_{\hat{n},\perp}=\Delta \mathbf{J} \cdot \hat{n}_{\perp}$ along $\hat{n}_{\perp}=\hat{n}\times\hat{y}$ axes as a function of angular distance from the center of the unstable point with $\hat{n}=\hat{x},\hat{z}$.
% $\Delta \mathbf{J}_{z}=\hat{z}\cdot \Delta \mathbf{J} $  for an initial Bloch vector  $\mathbf{J}_i \approx \{\epsilon_x, N/2 ,0\}$  as a function of $\epsilon_x $, and similarly $\Delta \mathbf{J}_{x}=\hat{x}\cdot \Delta \mathbf{J} $ for an initial Bloch vector $\mathbf{J}_i \approx \{0, N/2, \epsilon_z  \}$   as a function of $\epsilon_z$. 
% denoted as $\Delta \mathbf{J}{\hat{n},\perp}=\Delta \mathbf{J} \cdot \hat{n}{\perp}$, onto the axes $\hat{n}_{\perp}=\hat{n}\times\hat{y}$. 
In Fig.~\ref{fig3}\textbf{(C)}, we analyze the orthogonal projection of interaction-induced flows,
especially focus on the data points that initially displaced from the saddle point along $\hat{n}=\hat{x}$ and $\hat{z}$ axes (depicted by green and orange lines in Fig.~\ref{fig3}\textbf{(A)}), and calculate their projections along the $\hat{n} \times \hat{y}=\hat{z}$ and $-\hat{x}$ axes ($\Delta \mathbf{J}_z$ and $-\Delta \mathbf{J}_x$), respectively. The dynamics can be explained by noticing that when the Bloch vector is initially prepared in the $y$-$z$ plane, the effective mean-field magnetic field is along the $\hat{z}$ axis with a magnitude of $-4\chi_p {J}_z $. Conversely, when prepared in the $y$-$x$ plane, the field is along the $\hat{x}$ axis, with an amplitude of $4\chi_p  {J}_x $. Thus one expects these two perpendicular displacements to grow linearly in magnitude with initial Bloch vector displacement,  as we observed in Fig.~\ref{fig3}\textbf{(C)}.  

%\cl{Here, we focus on the short time dynamics to reveal the local flow lines and the exponential growth of TACT. The many-body energy gap emerging from the achieved TACT Hamiltonian can also protect the system from dephasing~\cite{Thompson2023MomentumExchange} allowing for a longer preservation of coherent dynamics.}

%------------------------------------------------------------------
\vspace{2mm}
\section{Two-axis counter-twisting with unstable points at north and south poles}
The original TACT Hamiltonian, as proposed by Kitagawa and Ueda \cite{KitagawaUedaOAT}, is defined as 
\begin{equation}
    \hat{H}^{\prime} = \chi \left( \hat{J}_+^2 + \hat{J}_-^2 \right) =2 \chi \left(\hat{J}_x^2 - \hat{J}_y^2 \right),
\label{eq:OriginalTACT}
\end{equation}

\begin{figure}[hbt!]
\centering
\includegraphics[width=0.5\textwidth]{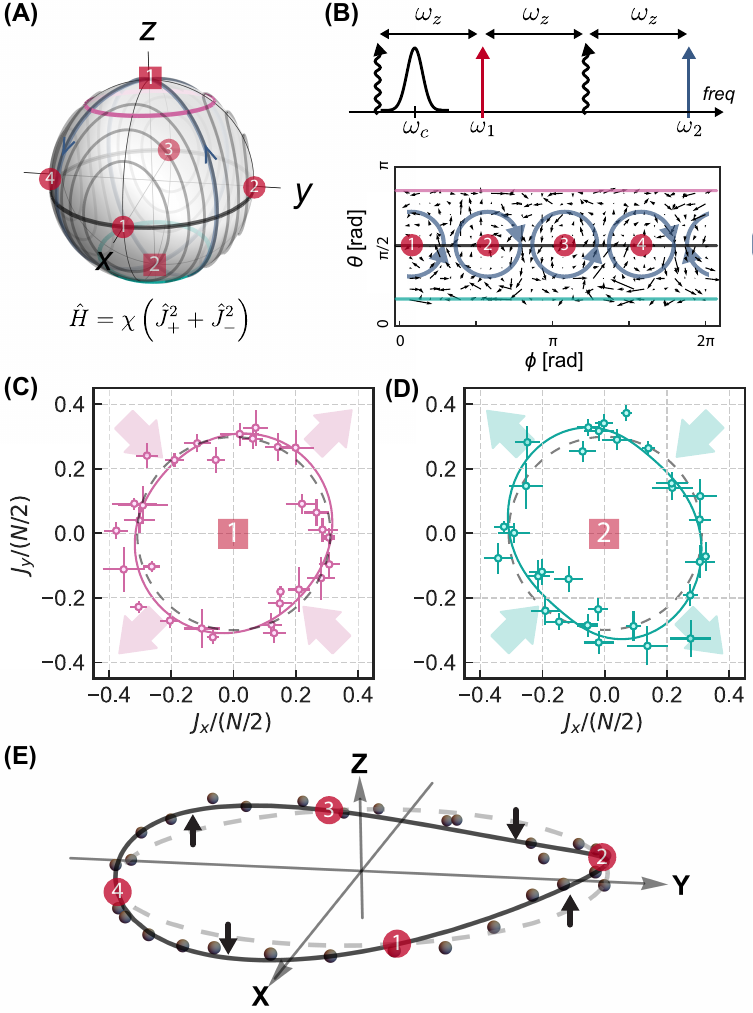}
\caption{\textbf{Two-axis counter-twisting with unstable points at north/south poles.}  
    \textbf{(A)} Flow lines  on the Bloch sphere. Three characteristic closed trajectories are displayed (green: $\theta_i=0.1\pi$, purple: $\theta_i=0.9\pi$, and black: $\theta_i=0.5\pi$ )
    \textbf{(B)} (top) Frequency diagram for the case in which both dressing lasers are positioned on the same side of  the cavity resonance.  (bottom) Equirectangular projection of the resulting dynamics (see text) 
    %targeted TACT Hamiltonian $\hat{H}^{\prime}$.
    \textbf{(C)}, \textbf{(D)} and \textbf{(E)}  Spin projections obtained after  time evolution of an initial distribution  of Bloch vectors with different polar angles  (green circle: $\theta_i=0.1\pi$, purple circle: $\theta_i=0.9\pi$  and black circle: $0.5\pi$ ). The deviation from the initial state distribution (gray dashed circles) is indicated  by the colored arrows.
 }  
\label{fig4}
\end{figure}

\noindent with the theoretical flow lines shown in Fig.~\ref{fig4}\textbf{(A)}. This Hamiltonian bears resemblance to the previously discussed TACT, with an additional $\pi/2$ rotation around the $y$-axis. It is characterized by unstable points at the north and south poles of the Bloch sphere, along with four stable points located on the equator at $\pm \hat{x}$ and $\pm \hat{y}$. 
%\clsout{In our setup, it seems to be impossible to choose the parameters canceling $\chi_e$ in Eq.~\eqref{eq:chi} at first glance \cl{when we consider only one modulation sideband from both dressing lasers with a common detuning $\Delta_{\bar{C}}$}. However, } 
We realize this Hamiltonian by placing the two detuned dressing lasers tones on the same side of the cavity resonance with $\Delta_{\bar{c}}=2\pi\times 700~\mathrm{kHz} > \omega_z$, such that one needs to account for another modulation sideband with opposite detuning from the cavity resoannce as shown in Fig.~\ref{fig4}\textbf{(B)} (top).  In this configuration, the previously ignored lower modulation sideband of the red dressing laser becomes significant. 
This sideband introduces an exchange interaction with the opposite sign of the other generated exchange interactions. By carefully selecting the detuning and the amplitude ratio of the two dressing laser tones, we can achieve a configuration where $\chi_e=0$, effectively only leaving the $\chi_p$ term (see supplement.) 
$\hat{H}^{\prime}$ offers the advantage of possibly being less sensitive to the presence of superradiance or collective decay for future squeezing generation in our system.

In Fig.~\ref{fig4}\textbf{(B)} (bottom), we show the measured flow lines for $\hat{H}'$ in the equirectangular projection. One can clearly identify four stable fixed points on the equator as expected.  To have a better intuition of the dynamics, instead of focusing on the dynamics on the whole Bloch sphere, we take a few cuts with initial Bloch vectors $\mathbf{J}_i$ prepared with $\theta_i=0.1\pi, 0.5\pi$ and $0.9\pi$ and study the dynamics separately. For the initial Bloch vectors prepared on the two circles with $\theta_i=0.1\pi$ and $0.9\pi$ (green and purple in Fig.~\ref{fig4}\textbf{(A)} and \textbf{(C)}/\textbf{(D)} near the north/south poles, the distribution of the states after the interaction $\mathbf{J}_f$ is plotted in Fig.~\ref{fig4}\textbf{(C)} and \textbf{(D)} (solid lines with fitted curves). The elliptical distributions with major axis orthogonal to each other explicitly show the squeezing and anti-squeezing axis near the north and south poles in this small displacement limit.  
In Fig.~\ref{fig4}\textbf{(E)}, the initial Bloch vector is prepared on the equator with different azimuthal angles (grey dashed line) and we plot $\mathbf{J}_f$ (black dot). The four zero crossings correspond to the four stable points on the equator. Between the stable points, the observed final states are deflected alternately above or below the equator as expected.

%-----------------------------------------------------
\vspace{2mm}
\section{Conclusion}

Here we have demonstrated the flexibility of our optical cavity simulator to engineer tunable XYZ Hamiltonians using two selected momentum states, without the need for Floquet engineering which in some cases might be challenging to allow access beyond mean-field dynamics in large many-body systems due to amplitude and phase noise on the Floquet control fields \cite{ThompsonPhaseNoise2012}.
The pair-raising/lowering processes are verified to be present for the first time by the resonant spectroscopic signal when the detuning $\delta$ is scanned as well as the direct cancellation of the exchange interactions that yielded dynamics of the Hamiltonian $\hat{H}'=\chi \left( \hat{J}_+^2 + \hat{J}_-^2 \right)$. By combining the correct relative balance of exchange and pair-raising/lowering contributions, we observed TACT dynamics for the first time.

In this work, we focus on the characterization of the Hamiltonian by probing the short-time dynamics at the mean-field level. In the future, it should be possible to go beyond mean field and create entangled states of interest for quantum simulation and metrology. Going beyond mean field will require reduced laser frequency noise on the dressing lasers that induce the interaction. 

Furthermore, by combining momentum states with internal states of the atoms or by adding more selected momentum states and dressing tones one should be able to engineer a toolbox for quantum state engineering, similar to what has been done with momentum states~\cite{meier2016observation,An2018,Meir2018,Gadway2021MomentumLattice, Lev2018SpinorSelfOrdering} . In our case however, in addition to the internal and external level control that tunes the synthetic dimensions,  we can further enjoy the rich opportunities offered by the tunable cavity-mediated interactions to engineer phenomena ranging from superfluidity and supersolidity~\cite{baumann2010dicke,Julian2017,leonard2017monitoring,schuster2020supersolid} to dynamical gauge fields and non-trivial topological behaviors~\cite{DynamicalPotential2013RMP,EsslingerSpinTexture2018,Esslinger2022DynamicalCurrent,Mivehvar2021,Ritsch2022Dynamical}. 

\begin{acknowledgments}
This material is based upon work supported by the U.S. Department of Energy, Office of Science, National Quantum Information Science Research Centers, Quantum Systems Accelerator. We acknowledge additional funding support from the VVFF, the NSF JILA-PFC PHY-2317149  and OMA-2016244 (QLCI), NIST, and DARPA/ARO W911NF-19-1-0210 and W911NF-16-1-0576, AFOSR grants FA9550-18-1-0319 and FA9550-19-1-0275. We acknowledge helpful discussions with Klaus Mølmer, Vladan Vuletić, and useful  feedback from Matteo Marinelli  and Calder Miller.
\end{acknowledgments}

% \clearpage
% \newpage

% \onecolumngrid

% \begin{center}
%     {\bfseries\Large{Appendix}}
% \end{center}
\appendix

\section{Derivation of Hamiltonian with two-color couplings}
We begin by considering the dispersive atom-cavity coupling Hamiltonian, with the exited state adiabatically eliminated as detailed in~\cite{Thompson2023MomentumExchange}. By applying two dressing lasers, each with frequencies $\omega_{1,2}$, amplitudes $\epsilon_{1,2}$, in  the rotating frame of the dressed cavity defined by $\hat{H}_c = \omega_c \hat{a}^\dagger \hat{a}$, the atom-cavity Hamiltonian can be written as:
\begin{equation}
\begin{split}
    \hat{H}_{0}=& \omega_{z}\hat{J}_{z}+\frac{g_{0}^{2}}{4\Delta_{a}}\hat{a}^{\dagger}\hat{a}\left(\hat{J}_{+}+\hat{J}_{-}\right) \\
    & + \epsilon_{1}e^{-i(\omega_{1}-\omega_c)t}\hat{a}^{\dagger}+\epsilon_{2}e^{-i(\omega_{2}-\omega_c)t}\hat{a}^{\dagger}+ \rm h.c.,
    \label{eq:eomac}
\end{split}
\end{equation}
with the atom-cavity detuning $\Delta_a=\omega_a-\omega_c$.
In addition, the cavity dissipation  can be modeled  by a Lindblad operator $\hat{L} =\sqrt{\kappa}\hat{a}$ with $\kappa$ denoting the cavity decay rate.

We can decompose the cavity field operator into $\hat{a}=\alpha(t)+\hat{b}$,
where $\alpha(t)$ represents an oscillating classical field. This field is described by the
\begin{equation}
\begin{split}
    \alpha(t) &=\alpha_{1}e^{-i(\omega_{1}-\omega_c)t}+\alpha_{2}e^{-i(\omega_{2}-\omega_c)t} \\
    \alpha_{1}&=\frac{\epsilon_{1}}{i\kappa/2+(\omega_{1}-\omega_c)}\quad\alpha_{2}=\frac{\epsilon_{2}}{i\kappa/2+(\omega_{2}-\omega_c)},
\end{split}
\end{equation}
where $\alpha_1$ and $\alpha_2$ are the coherence field amplitude inside the cavity  established  by the two dressing lasers.
We define the detuning for the two dressing lasers as $\Delta_{c1}=(\omega_{1}-\omega_c)+\omega_z$ and $\Delta_{c2}=(\omega_{2}-\omega_c)-\omega_z$.  Additionally, the average detuning and the differential detuning introduced in the main text can now be expressed as $\Delta_{\bar{c}}=\left(\Delta_{c1}+\Delta_{c2} \right)/2$ and $\delta = \Delta_{c2} - \Delta_{c1}$.
The resonance frequency diagram, depicted in Fig.~\ref{fig:setup}\textbf{(C)}, illustrates the case when $\Delta_{c1}=\Delta_{c2}=\Delta_{\bar{c}}<0$.

To derive the atom-only dynamics, we perform the standard second-order perturbation theory to eliminate quantum fluctuation $\hat{b}$, assuming the negligible excitation $\left\langle \hat{b}^\dagger \hat{b} \right\rangle \ll 1$. The effective Hamiltonian takes the form of
\begin{equation}
\begin{aligned}
&\hat{H}_{\rm eff} / \hbar 
  =\left(\frac{g_{0}^{2}}{4\Delta_{a}}\right)^{2}  \left[\frac{\Delta_{c1}|\alpha_1|^2}{\Delta_{c1}^{2}+(\kappa/2)^{2}} + \frac{\Delta_{c2}|\alpha_2|^2}{\Delta_{c2}^{2}+(\kappa/2)^{2}} \right] \hat{J}_{+}\hat{J}_{-}\\
+& \left(\frac{g_{0}^{2}}{4\Delta_{a}}\right)^{2} \frac{\alpha^*_2\alpha_1}{2}(\frac{1}{\Delta_{c1}+i\kappa/2}+\frac{1}{\Delta_{c2}-i\kappa/2})e^{i\delta t}\hat{J}_{+}\hat{J}_{+}\\
+& \left(\frac{g_{0}^{2}}{4\Delta_{a}}\right)^{2}\frac{\alpha^*_1\alpha_2}{2}(\frac{1}{\Delta_{c1}-i\kappa/2}+\frac{1}{\Delta_{c2}+i\kappa/2})e^{-i\delta t}\hat{J}_{-}\hat{J}_{-}
\end{aligned}
\end{equation}
Simultaneously, the dissipation is described by the effective Lindbard operator:
\begin{equation}
\hat{L}_{\rm eff}=\frac{g_{0}^{2}}{4\Delta_{a}}\sqrt{\kappa}[\frac{\alpha_1e^{-i\Delta_{c1} t} }{\Delta_{c1}+i\kappa/2}\hat{J}_{+}+\frac{\alpha_2e^{-i\Delta_{c2} t} }{\Delta_{c2}+i\kappa/2} \hat{J}_{-}].\label{diss}
\end{equation}
Here we have ignored two  additional  modulation sidebands that are detuned from the dressed cavity frequency by  $\Delta_{{c}1}-2\omega_z$ and $\Delta_{{c}2}+2\omega_z$\cite{Thompson2023MomentumExchange}, as their contribution to the dynamics is negligible.
For the dressing laser configuration in Fig.~\ref{fig4}\textbf{(B)}, one of the modulation sideband with detuning $\Delta_{{c}1}-2\omega_z \lesssim
\Delta_{\bar{c}}$ from the dressed cavity resonance becomes significant. This sideband introduces an additional exchange process into the effective Hamiltonian with an opposite sign to that of $\chi_e$. In order to  use that term to obtain the TACT Hamiltonian $\hat{H}^{\prime}$ by cancelling the exchange interaction term, we require:
\begin{equation}
    \frac{\Delta_{c1}|\alpha_1|^2}{\Delta_{c1}^{2}+(\kappa/2)^{2}} + \frac{\Delta_{c2}|\alpha_2|^2}{\Delta_{c2}^{2}+(\kappa/2)^{2}} =  \frac{\left(\Delta_{c1}-2\omega_z\right)|\alpha_1|^2}{\left(\Delta_{c1}-2\omega_z\right)^{2}+(\kappa/2)^{2}},
\end{equation}
which used for the measurements reported in Fig.~\ref{fig4}.

For the short time dynamics explored in Fig.~\ref{fig2} and Fig.~\ref{fig4}, we can ignore the superradiant dynamics given by Eq. \ref{diss} and only consider Hamiltonian evolution. In Fig.~\ref{fig3}, to study the slow dynamics near the unstable saddle point, both experimental data and theoretical simulation account for and subtract the effects due to superradiance with the same analysis as discussed below.

\section{Mean-field equation of motion and Stability analysis\label{secs2}}
In this section we study the fixed points and stability for the general XYZ model in Eq.~\eqref{eq:heff} and LMG model.
We start with the mean-field equations of motion for the  XYZ model in the Heisenberg picture,
\begin{equation}
    \begin{aligned}
\frac{d J_x }{dt}&=2\chi_y J_y J_z \\
\frac{d J_y }{dt}&=-2\chi_x J_x J_z \\
\frac{d J_z }{dt}&= 2\left(\chi_x - \chi_y \right)J_x J_y,
    \end{aligned}
\end{equation}
with $\chi_x = \chi_{e}+\chi_{p}$ and $\chi_y=\chi_e - \chi_p$ for our realization. 
The fixed points $\bf J_{\rm fix}$ for the aforementioned non-linear equations correspond to the spin aligned along the $\pm \hat{x}$, $\pm \hat{y}$ and $\pm \hat{z}$ directions. The torque is given by $\mathbf{T}(\mathbf{J})=\left(2 \chi_y J_y J_z, -2 \chi_x J_x J_z, 2\left( \chi_x - \chi_y \right) J_x J_y\right).$  The stability at the fixed point can be deduced from the Jacobian matrix,
\begin{equation}
    M(\mathbf{J})=\left(\begin{array}{ccc}
0 & 2 \chi_y J_z & 2 \chi_y J_y\\
-2\chi_x J_z & 0 & -2 \chi_x J_x\\
2\left(\chi_x - \chi_y \right) J_y & 2\left(\chi_x - \chi_y \right) J_x & 0
\end{array}\right).
\end{equation}

In the dressing laser configuration depicted in Fig.~\ref{fig:setup}\textbf{(C)}, we consider the condition where $\chi_x \leq \chi_y \leq 0$. Under this configuration, $\pm\hat{x}$ and $\pm\hat{z}$ represent the system's stable points, having eigenvalues of 
$\pm i2\sqrt{\chi_{x}\left(\chi_x - \chi_y\right)}$
and $ \pm i2 \sqrt{\chi_x \chi_y}$ respectively.
Conversely, $\pm \hat{y}$ are identified as the saddle points of the system, characterized by the real eigenvalues $\pm2\sqrt{\chi_y \left(\chi_{x}-\chi_{y}\right)}$. By fixing $\chi_x$, the system achieves its largest eigenvalue when $\chi_y = \chi_x /2$ thus $\chi_e=3\chi_p$, corresponds to the TACT Hamiltonian $\hat{H}= \chi_p \left( 2\hat{J}^2_x + \hat{J}^2_y \right)$.

Now we apply the same analysis  to the LMG model $\hat{H}=\chi \hat{J}^2_z + \delta \hat{J}_y $ with  equations of motion given by,
\begin{equation}
\begin{aligned}
    \frac{dJ_x}{dt} &= -2 \chi J_y J_z + \delta J_z \\
    \frac{dJ_y}{dt} &= 2 \chi J_x J_z \\
    \frac{dJ_z}{dt} &= - \delta J_x.
\end{aligned}
\end{equation}
Following the previously outlined procedure, we identify two stable points at $\mathbf{J}_{\rm fix} = \frac{N}{2} \left(0, \frac{\delta}{\chi N}, \pm \sqrt{1-\left(\frac{\delta}{\chi N}\right)^2} \right)$,
one stable point along $-\hat{y}$ direction and one saddle point along $\hat{y}$ direction, given $0<\delta < \chi N$. In the case when $\delta > \chi N$, the system exhibits two stable points along the $\pm \hat{y}$
directions, and  no saddle points. 
The eigenvalues of the Jacobian matrix at the saddle point are $\pm \chi N \sqrt{\frac{\delta}{\chi N} \left(1 -  \frac{\delta}{\chi N}\right)}$,  and for fixed $\chi N$, the maximum rate is achieved when $\frac{\delta}{\chi N} = 1/2$.

After examining the global flow lines, which reveal notable differences between the TACT and LMG models as previously discussed, our focus now shifts to the local flow lines near the saddle point along the $\hat{y}$ direction. We analyze this using the Holstein-Primakoff approximation, where the spin operators are represented as $\hat{J}_x \approx \sqrt{N} \hat{p}, \hat{J}_y \approx J - \hat{c}^\dagger \hat{c}, \hat{J}_z \approx \sqrt{N} \hat{x}$. Here, $\hat{c}^\dagger$ denotes a bosonic mode creation operator, and  $\hat{x}$ and $\hat{p}$  the corresponding position and momentum quadratures respectively. Consequently, the TACT Hamiltonian can be approximated by $\hat{H}= \chi \left(\hat{J}^2_x - \hat{J}^2_z \right)\approx - \chi N \left[(\hat{c}^\dagger)^2 + \hat{c}^2 \right]$. Similarly, for the LMG model (at $\frac{\delta}{\chi N}=1/2$) near the saddle point,  $H=\chi \hat{J}^2_z + \frac{\chi N}{2} \hat{J}_y \approx \chi N \hat{x}^2 - \frac{\chi N}{2} \hat{c}^\dagger \hat{c} \approx \chi N \left[(\hat{c}^\dagger)^2 + \hat{c}^2 \right]$, thereby locally resembling the TACT Hamiltonian.

\section{Four-photon spectroscopy}

\begin{figure}[hbt!]
\centering
\includegraphics[width=0.28\textwidth]{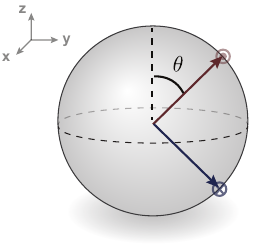}
\caption{Initial states for four-photon spectroscopy.}  
\label{figs1}
\end{figure}

In Fig.~\ref{fig:setup}E, we have shown a spectroscopic result of the four-photon resonance. For this experiment, we prepare two different initial Bloch vectors to be $\pi/4$ above or below the equator with projections nominally along $\hat{y}$, as shown in Fig.~\ref{figs1}. 

When on resonance with $\delta=0$, the ratio of the two dressing laser amplitudes is balanced to achieve an effective Hamiltonian in the form of $\hat{H} = \chi \hat{J}_x^2$, which preserve the azimuthal angle. 

The four-photon detuning $\delta$ can be intuitively understood as inducing a time-dependent one-axis twisting interaction $\hat{H}=\chi \hat{J}_{\phi\left(t\right)}^2$ where the twisting axis rotates as $\hat{J}_{\phi\left(t\right)} = \cos\left( \delta t /2 \right) \hat{J}_x + \sin\left( \delta t /2 \right) \hat{J}_y$. With the four-photon detuning $\delta\gg \chi N$, the effective Hamiltonian recovers to the exchange interaction $\hat{H} = \chi \left( \hat{\mathbf{J}} \cdot \hat{\mathbf{J}} - \hat{J}^2_z\right)$ which induces finite changes in the azimuthal angle of the initial Bloch vectors with the signs depending on the initial projection along $\hat{z}$. 

\begin{figure}[hbt!]
\centering
\includegraphics[width=0.45\textwidth]{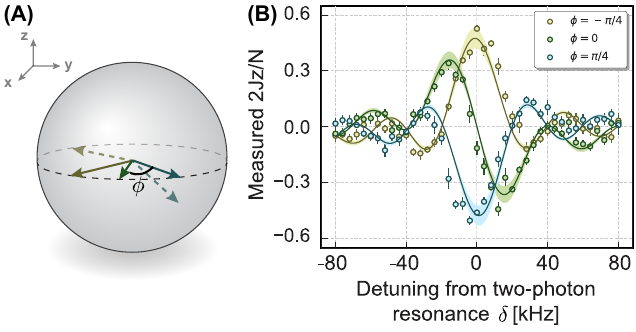}
\caption{Four-photon spectroscopy with initial states on the equator. (A) The initial Bloch vectors. (B) Interaction induced change in $J_z$ as a function of the four-photon detuning. }  
\label{figs3}
\end{figure}

With a different set of initial states, the four-photon resonance can also be witnessed by the change in polar angle or equivalently spin projection $J_z$. In this experiment, we prepare initial Bloch vectors on the equator but with different azimuthal angle $\phi=-\pi/4, 0$ and $+\pi/4$ as shown in Fig.~\ref{figs3}. 

When on resonance with $\delta=0$, the effective Hamiltonian $\hat{H} = \chi \hat{J}_x^2$ causes a positive/negative change in $J_z$ for initial states with $\phi=\pm \pi/4$. While for $\phi=0$, the initial Bloch vector is along $\hat{J}_x$ which commutes with the Hamiltonian and thus experience no change in $J_z$. With $\delta\gg \chi N$, $J_z$ in conserved under the effective exchange interaction $\hat{H} = \chi \left( \hat{\mathbf{J}} \cdot \hat{\mathbf{J}} - \hat{J}_z\right)$ for all three initials states.

\section{Sequences for flow vector measurement}

The local flow vectors are measured by the changes in polar angles $d\theta$ and azimuthal angles $d\phi$. Starting with all atoms in $\ket{\uparrow}$, the initial states are prepared by applying Bragg pulses with durations and phases determined by the initial state parameterized by $\theta_i$ and $\phi_i$. For all the Bragg pulses applied, the Rabi frequency is 8.3~kHz, leading to a $\pi$-pulse duration of 60. $\mu$s. Right after preparing the initial state, the interaction is applied before the wave packets separate. 

To measure the changes in polar angles, we apply a $\pi$-pulse along $\phi_i$, adjust the delay time to refocus the wave packets, apply a final $\left(\pi/2+\theta_i\right)$-pulse along the axis $\phi_i+180\deg$ to bring the Bloch vector nominally around to the equator, and measure the projection of the Bloch vector $J_z|_{d\theta}$ along $\hat{z}$. 

For measuring the changes in azimuthal angles, after the interaction, we apply a $\pi$-pulse around $\phi_i$, again adjust the delay time to refocus the wave packets, apply a final $\pi/2$-pulse around $\phi_i+90\deg$, and measure the projection of the Bloch vector $J_z|_{d\phi}$ along $\hat{z}$. 

With the two sequences described above, the changes in polar and azimuthal angle are mapped to the changes in the projection along $\hat{z}$, which can then be estimated by $d\theta = J_z|_{d\theta}/J$ and $d\phi=J_z|_{d\phi}/J$.

\section{Dynamics around the unstable saddle point}
% \begin{figure}[hbt!]
% \centering
% \includegraphics[width=0.6\textwidth]{figS2.pdf}
% \caption{Experimentally measured local flow (left) and numerical simulation result (right). Here, $\theta$ and $\phi$ are the polar and azimuthal angle of the initial Bloch vector.}
% \label{figs2}
% \end{figure}

In Fig.~\ref{fig3}, we measure the local flow vector as a function of the initial displacements from the unstable saddle point. For a more explicit comparison, the measured data (left) and the corresponding numerical simulation (right) results are shown below in Fig.~\ref{figs2}. The original data and the corresponding simulation are presented in the top row, and the result after subtracting superradiance are shown at the bottom. In both cases, we find a clear agreement between theory and experiment.

\begin{figure}[hbt!]
\centering
\includegraphics[width=0.45\textwidth]{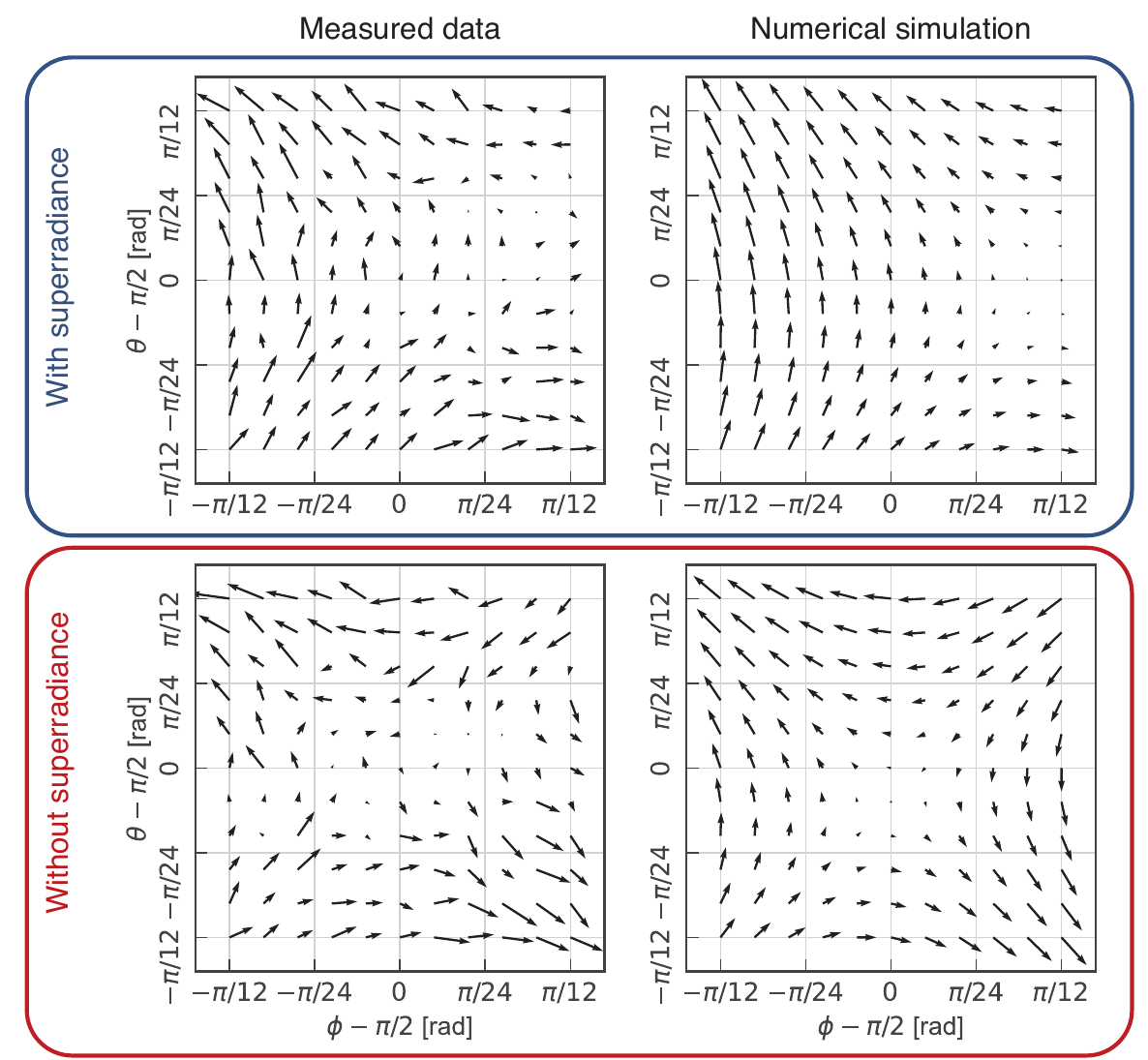}
\caption{Experimentally measured local flow (upper) and numerical simulation result (lower). We present the data before (right) and after (left) subtracting the superradiant, which both match with the theory.
Here, $\theta$ and $\phi$ are the polar and azimuthal angles of the initial Bloch vector.}
\label{figs2}
\end{figure}

For the measured data, we estimated the effect of superradiance by averaging the displacement along $\hat{z}$ for all data points. This common background displacement is then subtracted from all points.

For the numerical simulation, we consider the classical phase space structure for general XYZ models as discussed in Sec.~\ref{secs2} but with superradiance included. Now, the equations of motion become
\begin{equation}
\begin{aligned}
\frac{dJ_{x}}{dt} & =2\chi_{y}J_{y}J_{z}-\Gamma J_{x}J_{z}\\
\frac{dJ_{y}}{dt} & =-2\chi_{x}J_{x}J_{z}-\Gamma J_{y}J_{z}\\
\frac{dJ_{z}}{dt} & =2\left(\chi_{x}-\chi_{y}\right)J_{x}J_{y}+\Gamma\left(J_{x}^{2}+J_{y}^{2}\right).
\end{aligned}
\label{eq:superradianteom}
\end{equation}
Here, $\Gamma = \kappa \left(\frac{g_0^2}{4\Delta_a}\right)^2 \left( \frac{\left|\alpha_1\right|^2}{\Delta_{c1}^2+\kappa^2/4}-\frac{\left|\alpha_2\right|^2}{\Delta_{c2}^2+\kappa^2/4} \right)$ is the mean-field strength of the superradiance, which can be canceled by balancing the intra-cavity powers of the two dressing laser tones. The numerical simulation result of Eq.~\eqref{eq:superradianteom} is presented in the top right of Fig.~\ref{figs2}. The simulation result with superradiance subtracted is shown in the bottom right. The subtraction was performed following the same procedure of superradiance subtraction used in analyzing the experimental data.

\section{Initial state preparation}
Starting with about $10^8$ atoms in a MOT near the cavity center, we first apply polarization gradient cooling to load about $2\times 10^5$ atoms into the 813~nm red detuned optical lattice. After ramping down the lattice depth, $\Lambda$-enhanced grey molasses cooling is applied to reduce the ensemble temperature down to $6~\mathrm{\mu K}$. We then perform degenerate Raman sideband cooling for lowering the radial temperature down to less than $1~\mathrm{\mu K}$. An additional optical pumping is applied to transfer the atom into $\ket{F=2,m_F=0}$ state, which results in a radial temperature of $1.4(5)~\mathrm{\mu K}$ \cite{Thompson2022SAI}. 

About 1000 atoms are selected with RMS momentum spread $\Delta p<0.1 \hbar k$ using velocity-selective two-photon Raman transitions, giving a characteristic dephasing time of $150~\mu$s for two momentum state separated by $2\hbar k$. The remaining atoms are then removed with transverse radiation pressure \cite{Thompson2022SAI}. To enhance the interaction with the more favorable Clebsch-Gordan coefficients, the atoms are then transferred from $\ket{F=2,m_F=0}$ into $\ket{F=2,m_F=2}$ using series of microwave pulses \cite{Thompson2023MomentumExchange}. 

\section{Generating frequency tones for Bragg rotations and interactions}

Quantum non-demolition measurement and Bragg rotations of momentum states are realized by a single laser coupled into the cavity \cite{Thompson2023MomentumExchange}. Here, we refer to this laser as the atomic probe, which is stabilized to the blue of the atomic transition $\omega_{a}$ with detuning about $500$~MHz.

Driving Bragg rotations between $\ket{p_0-\hbar k}$ and $\ket{p_0+\hbar k}$ requires two different laser tones separated by $\omega_z$. These two tones are generated by first red shift the atomic probe frequency by 75~MHz with one acousto-optic modulator (AOM) and then blue shift it back with another AOM driven with two radio frequency (RF) tones at $\omega_\mathrm{RF1}=75$~MHz and $\omega_\mathrm{RF2}=(75-\omega_z/2\pi)$~MHz. The phase of the Bragg rotation $\phi_\mathrm{B} = \phi_\mathrm{RF2}-\phi_\mathrm{RF1}$ is defined by the relative RF phases of $\omega_\mathrm{RF2}$ and $\omega_\mathrm{RF1}$. To compensate the changing Doppler shift due to the free-falling, the frequency separation between the two tones are chirped by $25.11$~kHz/ms. 

For driving the interactions, we need two tones offset by $2 \omega_z$ with the separation chirped by $50.22$~kHz/ms. This is again realized by first red shift the atomic probe by 75~MHz with one AOM and then blue shift it back with another AOM. In this case, the second AOM is driven by two RF tones at $\omega_\mathrm{RF1}$ and $\omega_\mathrm{RF3}=(75-2\omega_z/2\pi)$~MHz. Here, $\omega_\mathrm{RF3}=2\omega_\mathrm{RF2}-\omega_\mathrm{RF1}$ is generated by first frequency doubling $\omega_\mathrm{RF2}$ then mixing with $\omega_\mathrm{RF1}$ to shift the frequency back down after proper frequency filtering. By doing so, we maintain the differential phase $\phi_\mathrm{int}=\frac{\phi_\mathrm{RF3}-\phi_\mathrm{RF1}}{2}-\phi_\mathrm{B}$ between the Bragg rotations and the interactions stabilized. This differential phase can also be rapidly controlled by the RF phase of $\omega_\mathrm{RF2}$.

% \section{Author contributions}
% C.L. and C.M. contributed to the building of the experiment, conducted the experiments and data analysis. J.K.T. conceived and supervised the experiments. H.Z. and A.C. contributed to the theoretical derivation and numerical simulations supervised by A.M.R. C.L., H.Z., A.M.R. and J.K.T. wrote the manuscript. All authors discussed the experiment implementation and results and contributed to the manuscript.

% \section{Data availability}
% All data obtained in the study are available from the corresponding author upon reasonable request.

% \section{Competing interests}
% The authors declare no competing interests.

% \bibliographystyle{apsrev4-2}
%\bibliographystyle{arthur}
\bibliography{Hamiltonian_Engineering.bib}{}

\end{document}